\documentclass[prb,preprint]{revtex4-1} 

\usepackage{amsmath} 

\usepackage{graphicx} 

\begin{document}

\title{The Process of Transforming an Advanced Lab Course: Goals, Curriculum, and Assessments}
\author{Benjamin M. Zwickl}
\affiliation{Department of Physics, University of Colorado Boulder, Boulder, CO 80309}
\email{benjamin.zwickl@colorado.edu} 
\author{Noah Finkelstein}
\affiliation{Department of Physics, University of Colorado Boulder, Boulder, CO 80309}
\author{H. J. Lewandowski}
\altaffiliation[Also at ]{JILA, University of Colorado Boulder, Boulder, CO 80309} 
\affiliation{Department of Physics, University of Colorado Boulder, Boulder, CO 80309}
\date{\today}

\begin{abstract}
A thoughtful approach to designing and improving labs, particularly at the advanced level, is critical for the effective preparation of physics majors for professional work in industry or graduate school.  With that in mind, physics education researchers in partnership with the physics faculty at the University of Colorado Boulder have overhauled the senior-level Advanced Physics Lab course.  The transformation followed a three part process of establishing learning goals, designing curricula that align with the goals, and assessment.  Similar efforts have been carried out in physics lecture courses at the University of Colorado Boulder, but this is the first systematic research-based revision of one of our laboratory courses.  The outcomes of this effort include a set of learning goals, a suite of new lab-skill activities and transformed optics labs, and a set of assessments specifically tailored for a laboratory environment.  While the particular selection of advanced lab experiments varies widely between institutions, the overall transformation process, the learning goals, and the assessments are broadly applicable to the instructional lab community.
\end{abstract}

\maketitle

\section{Introduction}

Labs at the undergraduate level, especially senior-level advanced lab courses, are expensive to equip and impact few students in comparison to larger lecture courses.  Financially difficult times may prompt some institutions to reconsider whether such courses should be supported in the future.\cite{Chodos2007}  The advanced lab is an effective educational environment and we propose a model of lab course transformation that focuses on outcomes of student learning while still honoring the existing structure at a particular institution.  We present both the process and materials to support other institutions in systematically modifying laboratory courses to maximize learning through a process that encourages defining clear learning goals, aligning curriculum with those goals, and assessing students' learning.

There is a national need to make excellent undergraduate laboratory courses.  Uninspiring introductory laboratory courses have been cited as a significant reason for low retention of science, technology, engineering, and mathematics (STEM) majors.\cite{PCAST2012}  In physics, where introductory courses serve many more non-majors than majors, it is a particular concern that these courses inspire and motivate all students, which includes students in a range of STEM disciplines.  Introductory lab courses that focus on research and design have been shown to be effective at increasing STEM retention for all students, including those in populations traditionally underrepresented in STEM fields.\cite{Fortenberry2007}  Lab courses that have carefully considered the goals and methods of instruction can have significant impacts in influencing students' attitudes and scientific practices.\cite{Etkina2006b, Etkina2010}  It is time to consider a research-based redesign at the advanced level, establishing our laboratory goals for the physics major and working backwards through the sequence of lab courses to create a comprehensive curriculum in experimental physics.

In addition to these general reasons for improving lab instruction, the transformation of the advanced lab course at the University of Colorado Boulder (CU) was motivated by additional local concerns, which may also be familiar to others.  These included (1) a disorganized array of equipment, some of which was out-of-date and non-functional, (2) a lack of clear goals that resulted in an array of disconnected experiences that did little to emphasize certain basic experimental skills, (3) indications that the course negatively affected students' attitudes about the nature of experimental physics, (4) indications that some students felt their lack of success in the lab course reflected on their personal abilities, and (5) faculty sentiment that the course was difficult to teach, particularly for faculty teaching the course for the first time.  

With the challenges of the existing advanced lab course in mind, the goals of the laboratory transformation are clear.  First, the course should be an excellent learning experience for students.  It should prepare students for research by integrating them into the practices and community of experimental physicists.  Second, the course should be an excellent teaching experience for faculty by providing the resources and structure so that faculty can concentrate on having quality interactions with students by sharing their expertise in experimental physics.  Third, all aspect of the course should be excellent, which includes the physical space and equipment, the organizational structure and schedule for the course, and the curriculum and pedagogy.  Finally, the project should be documented and should expand the physics education research (PER) knowledge base on lab courses so that other instructors and institutions can systematically build on these efforts.

One of the principles that guided our redesign was to honor as many of the existing institutional structures as possible and only make changes that are needed to achieve our goals.  To that end, it is worth documenting the nature of the pre-transformed course, since some of the outcomes of the project are directly tied to retaining these features.  First, the previous course, despite its limitations, had a strong foundation of sophisticated apparatus and significant faculty involvement; many students regarded it as the best in the undergraduate lab course sequence.  In terms of structure, the CU Advanced Lab course is in many ways typical of similar courses in other colleges and universities.  The typical enrollment in the advanced lab is 20-30 students per year in a department with approximately 370 undergraduate majors.  Not every student completes the course because the requirement can also be fulfilled by an undergraduate research experience.  The lab course is not specifically aligned with any of the canonical undergraduate lecture courses.  The content covers a wide range of topics including nuclear physics (gamma ray spectroscopy), particle physics (cosmic ray muon lifetime), condensed matter physics (scanning tunneling microscope, NMR), AMO physics (saturated absorption spectroscopy, magneto-optical trapping), nonlinear dynamics (soliton propagation in a nonlinear waveguide), acoustics (acoustic network analyzer), and physical optics (diffraction, polarization, interferometry).  The official course schedule consisted of three lab hours and two lecture hours per week, though students spent 10-15 hours in lab per week and had around-the-clock access to the lab space.  The lectures typically covered topics in experimental physics, but were not directly applicable to the lab course.  The course had 10 weeks of guided labs, each of which took 1-3 weeks to complete, allowing students to complete 4-6 different experiments.  The final five weeks of the course were devoted to a final project, which could be an extension of an existing lab or some other topic depending on students' interests.

The lab course is a special instructional environment and we seek to build on it's unique aspects (while realizing it's limitations).  These unique aspects include: a classroom environment ready for group work and frequent instructor-student interactions, which are hallmarks of active engagement; significant investments of dedicated space and resources; low student-teacher ratios; and finally, the courses are usually taught by expert practitioners of experimental physics.  Despite these assets, typically a student works in lab 10-15 hours per week over 15 weeks, which cannot duplicate the depth of a more substantial undergraduate research experience.  Thus, the course should focus on meeting important learning goals that can be addressed in shorter periods of time.  Similarly, the sophistication and expense of the equipment cannot equal a research lab that is funded by external grant funding.  Such highly specialized physics content and apparatus also make the course more difficult to teach for instructors from a different sub-field of physics.  Our lab redesign emphasizes equipment than can be adapted to a variety of experimental investigations.  The experiments were chosen not because of their uniqueness in the laboratory, but for their ubiquitousness.  A goal of the project was to reuse and adapt as much of the existing lab equipment as possible.  In a time of financial insecurity in higher education, we must continue to use resources wisely and convincingly demonstrate the value of the advanced lab. 

\section{Prior literature on lab transformation}

Compared to lecture courses, there is little existing research-based advice on how to transform a lab course.  Introductory laboratory courses have received the most attention and research on upper-division lab courses is sparse.  Laboratory instruction for the preparation of physicists and teachers of physics has been a part of college science instruction for over a century in the United States.\cite{Hall1892}  Particular attention has been given to developing and sharing individual experiments.  For instance, the American Journal of Physics (AJP) continues to feature an undergraduate-level experiment in nearly every issue.  Additionally, an organization of physics lab instructors, the Advanced Laboratory Physics Association (ALPhA, http://advlab.org), is training other instructors to conduct new experiments through hands-on ``Immersions.''   Even with all these effective resources, there has been relatively little attention given to pedagogy and how to integrate these activities into a course or part of a multi-year lab curriculum.  

There are some examples of the development of entire laboratory courses.  The first documented in AJP is Harvard's course in atomic physics, published in 1934.\cite{Oldenberg1934}  More recently at the introductory level, the Investigative Science Learning Environment at Rutgers \cite{Etkina2007a} is a combined lecture-lab environment that focuses on developing students' scientific practices such as experimental design.\cite{Etkina2006b}  A combined lecture-lab Studio approach has been applied to the upper-division optics course at Kansas State University.\cite{Sorensen2011}  Other more traditional upper-division lab courses have been documented that focus on optical spectroscopy,\cite{Blue2010} laser physics,\cite{Henningsen2011} and a series of single-photon quantum optics experiments.\cite{Galvez2005,Pearson2010}  It is interesting to note that many of these upper-division transformed courses involve optics and use a common set of basic experimental tools that can be reconfigured to perform a variety of experiments.  

Our effort places a heavy emphasis on the goals, curriculum, and assessment that is rarely documented in other laboratory experiences.  The work presented here is focused on lab pedagogy, not apparatus and theoretical background.  

\section{Model for Course Transformation}
Overhauling a lab course is a complex task.  There are a variety of important components, such as the equipment and organization, that need to be attended to in addition to pedagogy, assessment, and curricular development. In order to remain focused on student learning as the ultimate goal, it helps to have a process for course transformation.  

\begin{figure}[h!]
\centering
\includegraphics[width=3.0 in]{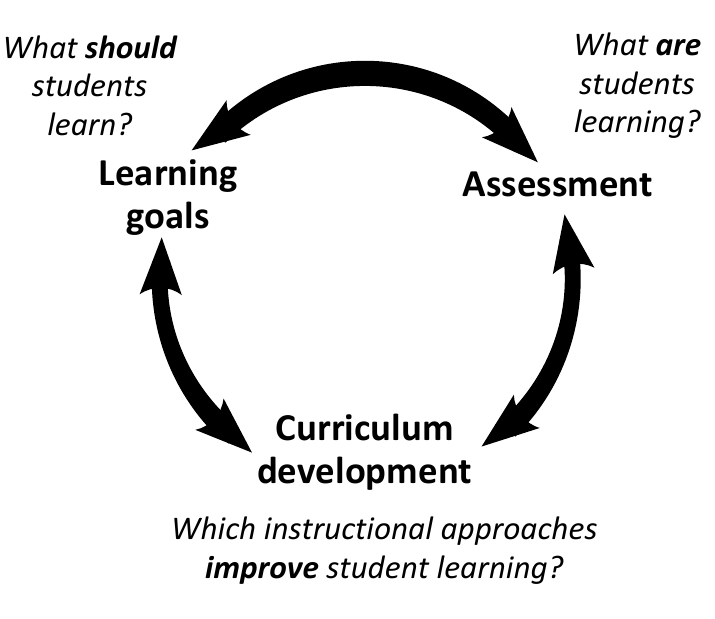}
\caption{The three coordinating aspects of the CU Science Education Initiative course transformation model.}
\label{fig:SEI_model}
\end{figure}

The University of Colorado Boulder Department of Physics has made significant changes to upper-division physics courses such as Classical Mechanics,\cite{Pollock2011a} Electricity \& Magnetism,\cite{Pepper2012} Modern Physics,\cite{McKagan2008a,	Baily2010} and Quantum Mechanics.\cite{Goldhaber2009}  These course transformations have used a process promoted by the CU Science Education Initiative (SEI) that involves three iterative phases: defining learning goals, developing curriculum that align with the goals, and assessment.\cite{Wieman2010, Chasteen2011}  Fig.~\ref{fig:SEI_model} rephrases these three elements as questions: (1) What should students know? (2) Which instructional approaches improve student learning? and (3) What are students learning?  

One of the benefits of this model is that sustainability is built into the process.  Faculty are consulted at the earliest stages of the project and all changes to the course structure and curriculum  align with consensus course goals and are further refined based on the assessments.  At CU, the transformation process has usually been coordinated by post-doctoral researchers in physics education research, but it could be equally accomplished by any faculty member.  As the transformed courses are taught, feedback is sought from faculty and students to make continual improvements to the course.

In late 2010, with support from the National Science Foundation, we began a comprehensive transformation of the advanced lab course using the model described above.  Classroom observations and discussions with faculty began in Spring 2011; learning goals were established by April 2011; curriculum was developed and space was renovated throughout Summer and Fall 2011; the first implementation of the redesigned course was taught in Spring 2012.  All of the assessment data were taken during the Spring 2012 semester, when 21 upper-division Physics and Engineering Physics majors enrolled in the course. 

\section{Step 1: Learning Goals}

The first step in the SEI Transformation Model is to establish consensus learning goals among the faculty.  Involving a large number of faculty at the beginning of the transformation process ensures any changes are not imposed on the faculty, but rather are directed by the collective voice of the department.  All CU physics faculty were invited to be interviewed about learning goals for the advanced lab.  Fifteen faculty responded and were interviewed for about one hour, which gave each faculty member a significant amount of time to share their personal views about the goals and purposes of lab courses.  Faculty were presented with a mix of unguided and guided questions such as ``What is the purpose of a good lab course?''  or ``What is the goal of communication in the course?'' or ``What abilities do you look for when hiring an undergraduate researcher or a new graduate student?''
 
After the faculty interviews were completed, the goals were grouped into three categories: consensus structure goals, consensus learning goals, and learning goals that warranted further discussion.  The consensus structure goals included topics like making the course easier to teach and providing support for advising students during the final project portion of the course.  The consensus learning goals included topics like data analysis and measurement automation using LabVIEW.  Finally, some learning goals elicited a variety of opinions from faculty, such as the appropriate balance of oral and written reports and goals for lab notebook usage. 

A summary of the interviews was presented to the faculty at two working group meetings.  The faculty present quickly affirmed the value of the consensus learning goals.  The majority of the discussion was devoted to the learning goals that warranted further discussion.

Altogether, 21 faculty (66 total faculty in physics) participated in the interviews  and meetings.  This group included a range of sub-disciplines including AMO physics (8), condensed matter (3), high energy particle (3), plasma (2), PER (3), and mechanical engineering (2).  Since the nature of experimental physics can vary between the sub-disciplines it was helpful to have such a diverse response from the department.  

\begin{figure}[h!]
\centering
\includegraphics[width=0.45\textwidth, clip, trim=0mm 0mm 0mm 0mm]{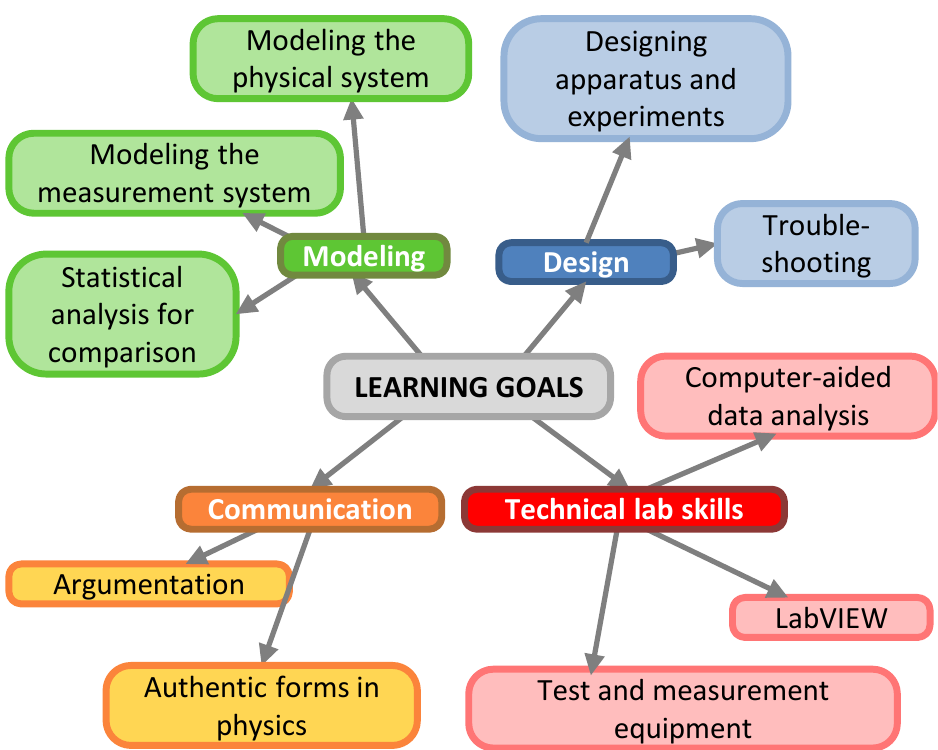}
\caption{Learning goals for the advanced physics lab course.  The full detailed learning goals document can be accessed at http://tinyurl.com/Advanced-Lab-LGs.}
\label{fig:Learning_goals}
\end{figure}

\begin{figure*}[h!]
\centering
\includegraphics[width=1.0\textwidth, clip, trim=0mm 0mm 0mm 0mm]{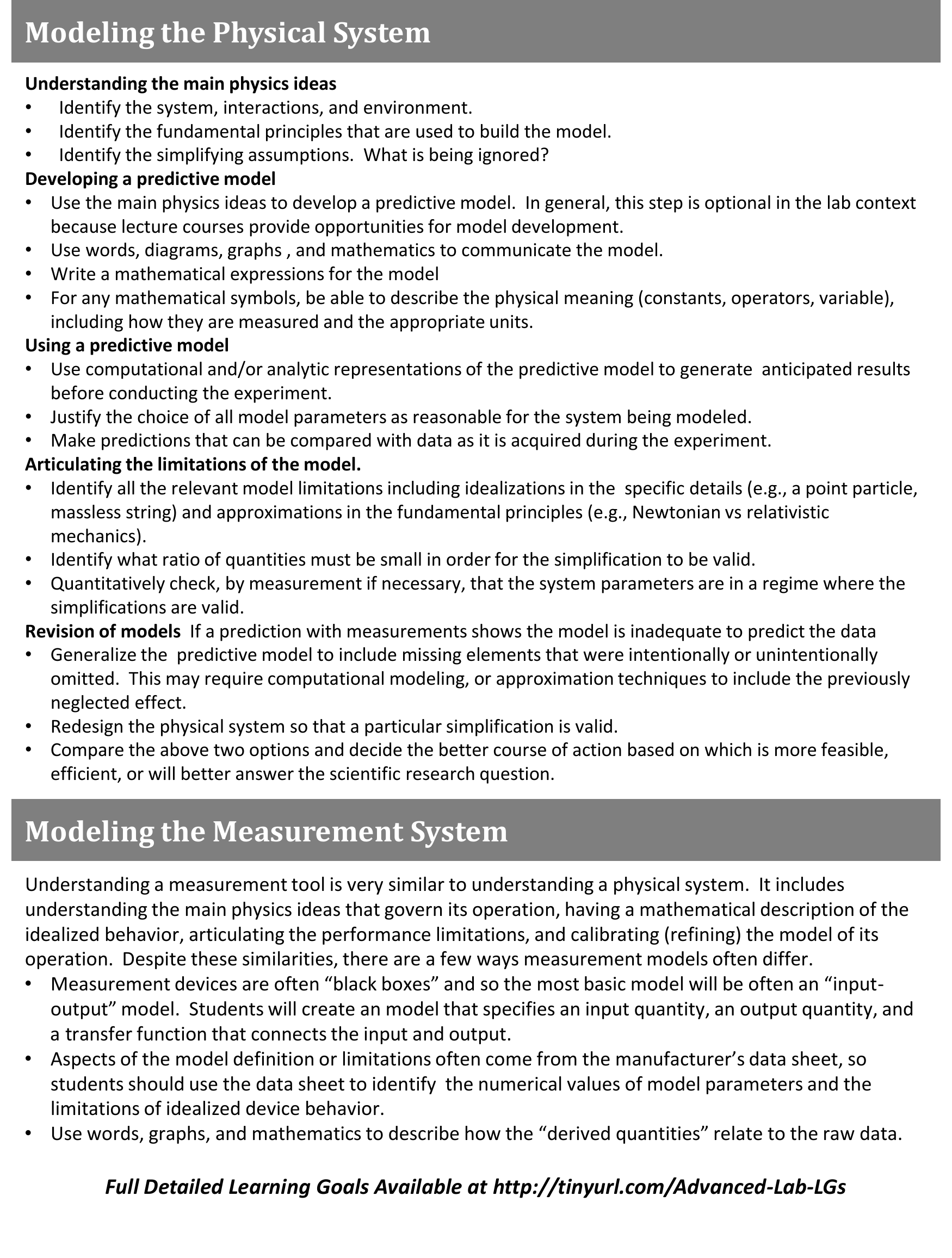}
\caption{Partial listing of the detailed learning goals for the advanced physics lab course.  The full detailed learning goals document can be accessed at http://tinyurl.com/Advanced-Lab-LGs.}
\label{fig:Learning_goals_detailed}
\end{figure*}

After much discussion, comparison with labs at other institutions, reference to the literature on physics education research, and a semester of classroom observations, four categories of learning goals emerged: modeling, design, communication, and technical lab skills.  Fig.~\ref{fig:Learning_goals} presents a summary of these goals with one more level of detail. However, learning goals are only useful when they are associated with specific observable and measurable outcomes.   An overview of the learning goals is below.  Fig.~\ref{fig:Learning_goals_detailed} shows a portion of the detailed learning goals, and the full document can be accessed online.\footnote{{http://tinyurl.com/Advanced-Lab-LGs}}    It is interesting to note that faculty did not emphasize goals centered around specific physics content or lab techniques.  Instead, faculty made it clear that the primary attribute of a good experiment is that students use a deep knowledge of physics and develop expert-like habits of mind while carrying out experiments and using measurement tools.

\subsection{Modeling}
Models can be defined as simplified abstract versions of real objects and their interaction, which have predictive power and a specified limited range of applicability.\cite{Etkina2006}  The process of modeling includes the development, use, testing, and refinement of models.  Centering physics instruction around the activity of modeling was proposed early on by David Hestenes\cite{Hestenes1987} and developed more fully through the Modeling Instruction program, which is used within high school and introductory college classrooms.\cite{Wells1995, Brewe2008}  One immediate appeal of a model-centered approach is the belief that it represents an authentic description of how professional researchers do physics.   A similar conclusion was reached in discussion with faculty.  According to faculty, the best experiments involve quantitative measurements of interesting physical systems and those measurements should be compared with predictions from theoretical models.  This explains why the traditional physics labs have emphasized error analysis, which deals with the comparison between data and predictions. However, by encapsulating error analysis within a larger context of modeling there is the opportunity to focus students' attention beyond statistics and onto a greater understanding of the physical ideas.  Modeling includes identifying the idealizations and approximations in the models of the physical system and measurement tools, so deviations from these idealizations are now the sources of systematic error.  Students go beyond suggesting sources of error to quantitatively including systematic effects in their model and/or refining their apparatus to better match the idealizations.  In this way, modeling provides a natural way to introduce systematic error into a broader discussion of measurement and uncertainty.   The key is that a model is not just an equation, but also includes a set of simplifying assumptions and idealizations that limit the applicability of the model.

There are three main components within the modeling category: modeling the physical system, modeling the measurement system, and using statistical analysis to compare the data and predictions.  Modeling the physical system means understanding the basic physics ideas that go into a predictive model, understanding the limitations of the model, using the model to make predictions, and revising the model based on experimental results.  Modeling the measurement system is similar, but involves understanding how the physical phenomena is connected to the data through the measurement tools.  The statistical analysis component involves many of the traditional techniques of data analysis and error analysis common in undergraduate physics labs.

\subsection{Design}
Design encompasses the activities of designing and redesigning apparatus and experiments and troubleshooting those systems.  These learning goals, include activities such as creating testable research questions, wisely designing experiments based on goals needed to convincingly answer the research questions, calibrating measurement tools, and continually reflecting on the results.  

Troubleshooting should be viewed as a condensed version of the scientific process.  Students should bring to bear all their skills in modeling and experimental design while troubleshooting.  In particular, they should make predictions about what each part of the apparatus should do, measure what it actually does, decide if the difference is significant enough to investigate further and/or identify a particular component as the source of discrepancy and then fix it.

\subsection{Communication}
Our faculty agreed that good scientific communication skills are some of the most valuable skills learned as an undergraduate.  Communication was divided into two main components.  The first is argumentation, which is the process of convincing an audience of a claim using evidence and reasoning and by considering other alternative claims as less well-supported.  Second, we want students to be able to present those arguments in forms that are authentic to physics research, which includes oral presentations and written papers.  While posters are also an authentic form of discourse, we did not include poster presentations in our learning goals.

\subsection{Technical lab skills}
Our faculty wanted to develop students' expertise in the most widely applicable lab skills and they placed a lesser emphasis on specific instrumentation and techniques.  These general purpose lab skills included using basic test and measurement equipment like oscilloscopes, using LabVIEW for measurement and automation, and using standard software packages like Mathematica for data analysis.  LabVIEW was not previously taught in the course, but was added because faculty cited it as one of the most useful skills for new graduate students to have when starting in a research lab.  We also placed a significant emphasis on computer-aided data analysis because it was not taught adequately in earlier courses and we found many senior-level students lacked analysis skills faculty desire for students.

\section{Step 2: Curriculum and Course Redesign}

Once the learning goals were established, all aspects of the course structure and curriculum were evaluated and some were transformed as necessary to align with the goals.  

\textbf{Tools.} The major changes to the equipment included the implementation of standardized optics workstations that can be adapted to any of the optics experiments conducted in the lab.  The individual optics and optomechanical components are kept in a central location, similar to research labs. Each workstation is furnished with a set of test and measurement equipment including a DC power supply, oscilloscope, and waveform generator.  All of the newly purchased optical, data acquisition, and test and measurement equipment is general purpose, flexible-use, and low-cost lab grade.

\textbf{Space.} In order to get students to participate in science (i.e., model, design, communicate, and use tools) we wanted to create an engaging space that had the appearance, organization, and functionality of a research lab.  In addition, we wanted to make the space easy to use and maintain because faculty rotate into the course and need to quickly become familiar with lab apparatus.  Major changes to the space included the conversion of an unused office space into a collaborative classroom space, which is adjacent to the lab space.  This space also serves as a venue for in-class student presentations. Also, any unused and/or broken equipment was removed from the lab space to simplify the organization.

\textbf{Time.} A two-hour-per-week ``lecture''  portion of the course was rescheduled to meet in the collaborative classroom space immediately before each lab section--providing continuity in both space and time with the weekly experimental activities.  For use during the lecture times, we developed a suite of short lab-skill activities on topics that students commonly struggled with and that would be valuable for the majority of the labs carried out during the semester.  We emphasized uncertainty, computer-aided data analysis, and LabVIEW during these collaborative work times.  In order to provide on-demand support for students and to ease adoption by faculty, a series of screencasts were developed and posted on YouTube demonstrating common activities like importing data into Mathematica (\url{http://www.youtube.com/compphysatcu}).  The lecture time was also used for students to give oral presentations to their classmates and instructors.

\textbf{Curriculum.} Finally, the learning goals were applied to the redesign of individual lab activities.  Generally speaking, a well-designed activity should, when completed by students, demonstrate their proficiency in one or more of the established learning goals.  The physics content and apparatus from an existing series of optics labs were retained, while the lab guides were modified to align with the learning goals.  Not only was this more cost effective than the development of new apparatus, but it allowed us to focus on the development of scientific process skills such as modeling and design.

\subsection{Example: New Gaussian Laser Beams Lab}
It is a general rule of teaching that the first class of the semester establishes students' expectations for the remainder of the semester.  As such, it was important that the first lab of the semester should directly emphasize the learning goals we hope to achieve through the course.  Rather than follow a set of prescriptive activities, students are charged with completing challenges that included technical lab skills such as aligning a laser beam through a randomly oriented tube using two mirrors, data analysis such as nonlinear least-squares fitting to extract the Gaussian beam width of the laser, using LabVIEW to efficiently acquire data, testing a model of Gaussian laser beam propagation and comparing with predictions from the ray theory, and developing a model of the amplified photodetector based on the manufacturers specifications and testing the specifications through a calibration experiment.  While these activities may seem basic, they are actually challenging activities that develop practices essential to successful research in experimental physics.

\subsection{Example: Modified Polarization Lab}
The previous version of our polarization lab included a demonstration of Malus' Law and of circularly polarized light produced by a quarter-wave plate.  The biggest limitation of the previous lab was not in the apparatus, but in the level of understanding the lab promoted.  In the original framing, students were given the equation for Malus' law $I_\textrm{trans}=I_\textrm{inc}\cos^2\theta$, and a qualitative description of the quarter-wave plate, but did not have access to a model of polarized light or polarization optics in terms of electric fields.  Students had very little ability to make any qualitative or quantitative predictions about a series of polarization filters and wave plates.  In addition, without a predictive model of the optical system, it was not possible to discuss easily observable systematic error effects.  As a first step in meeting our learning goals of modeling the physical system and measurement apparatus, students now develop a model of polarized light using Jones matrix formalism.\cite{Jones1941}  One benefit of the more sophisticated predictive model is that the measurements become more sophisticated as well.  For example, the Jones formalism model can be used to fit data and quantitatively determine the parameters of elliptical polarization.  One group of students further extended the techniques and models by performing an ellipsometry experiment for their final project.  The students used ellipsometry to determine the thickness and index of a unknown thin film on a silicon substrate--a measurement frequently performed in industrial lab settings.  In the revised polarization lab, the equipment is identical to the previous lab, but the focus is now on modeling the experimental system, predicting results, comparing with data, and uncovering and modeling systematic error effects.

\section{Step 3: Assessment}

In order to continually improve the course, it is essential to assess what effect the course is having on students' learning.  The assessment takes place in four ways.  First, the activities themselves are meant to be assessments of our learning goals.  Successful completion of an activity aligned to the learning goal is both the mechanism for developing proficiency\emph{and} for demonstrating proficiency.  This approach assesses the widest variety of our learning goals.  In addition we employed three surveys: (1)  weekly surveys asking students to evaluate their lab experience, (2) a self-assessment of learning gains survey given at the end of the semester, and (3) the Colorado Learning Attitudes about Science Survey for Experimental Physics (E-CLASS).

\subsection{Weekly Surveys}
The weekly surveys provide a record of students attitudes about individual labs. Students were asked to reflect on the past week and tell what they learned about the most, what they want to learn more about, and what they would recommend changing about the past week's activities.  Sometimes students mentioned typos in the lab guide, but more often they gave general comments about parts of the activities that were particularly time-consuming, confusing, or difficult.  The weekly surveys also helped us document a couple of instances of inadequate equipment (e.g., beam splitters, cameras for imaging interference patterns).  Other valuable feedback noted was that while 83\% of students thought the Mathematica and LabVIEW activities were very helpful for learning, the time allotted for completion was too short, which students found frustrating (only 65\% said they enjoyed the activities).  In future semesters, we will spread the activities over a larger number of days.

\subsection{Self-Assessment of Learning Gains (SALG)}
The Self-Assessment of Learning Gains\cite{Seymour2000} (\url{http://www.salgsite.org}) is an online survey tool that asks students to assess how different aspects of a course promote students' learning.  The SALG can be customized by the instructor to align it with the particular activities, content areas, and skills emphasized in a course.  By focusing on learning gains, the survey is intended to shift students' evaluations from how much they liked the course to how much they learned from it.  Students have been shown to self-evaluate accurately enough to make the survey beneficial for documenting the impact of course modifications.\cite{Seymour2000}  Additional evidence for the usefulness of the SALG in lab courses has been its extensive use in evaluating undergraduate research programs like the NSF Research Experience for Undergraduates (REU) program.  The SALG version known as the Undergraduate Research Student Self-Assessment (URSSA) has documented gains in skills, knowledge, confidence and identity as a scientist, and clarity in career goals.\cite{Hunter2009}  For our customized SALG survey, one instance of helpful feedback was that the course grading guidelines and student feedback on written reports were not rated as very helpful in promoting learning. This feedback is leading to the development and adoption of rubrics for written reports, oral presentations, and final projects to clarify the expectations and standardize and speed up the grading process.

\subsection{E-CLASS}
The last form of assessment given to the students was the recently developed Colorado Learning Attitudes about Science Survey for Experimental Physics (E-CLASS).\cite{Zwickl2012}  Whereas the SALG focuses on learning gains, the E-CLASS, in the tradition of the earlier CLASS survey, characterizes students' attitudes about learning physics.\cite{Adams2006}  The E-CLASS focuses specifically on assessing attitudes about habits of mind and practices in experimental physics and compares these with the grading requirements and common practices in their recently completed lab course.  For example, Table\ \ref{tab:GradeEmphasis} shows the degree to which students thought particular practices were important for earning a good grade in the Advanced Lab course.  Looking at those practices ranking near the bottom, it is no surprise that reading scientific journal articles and thinking up questions to investigate rank as only moderately important for earning a good grade.  This lower emphasis arises because the guided labs conducted during the first 10 weeks do not draw on the wider scientific literature nor let students pose their own.  Nevertheless, students said that thinking about the purpose of the instructions was only moderately important for earning a good grade, which is an indicator that students are able to complete the activities without a great deal of reflection on the overall goals of what they are doing.  In response to this finding, we hope to incorporate metacognitive strategies that encourage more reflection during lab, as has been done in some other transformed lab curricula.\cite{Tien1999, LippmannKung2007, Etkina2010} 

\begin{table*}[t]
\begin{tabular}{p{0.9\textwidth}  p{0.08\textwidth}}

``How important for earning a good grade in this class was...''  & Mean\\ 
``...using a computer for plotting and analyzing data?'' & 4.94\\ 
``...writing a lab report that made conclusions based on data using scientific reasoning?'' & 4.69\\ 
``...understanding how the experimental setup works?'' & 4.63\\ 
``...learning to use a new piece of laboratory equipment?'' & 4.63\\ 
``...understanding the equations and physics ideas that describe the system I am investigating?'' & 4.56\\ 
``...communicating scientific results to peers?'' & 4.50\\ 
``...using error analysis (such as calculating the propagated error) to better understand my results?'' & 4.50\\ 
``...making predictions to see if my results are reasonable?'' & 4.31\\ 
``...working in a group?'' & 4.25\\ 
``...understanding how the measurement tools and sensors work?'' & 4.13\\ 
``...understanding the performance limitations of the measurement tools?'' & 4.06\\ 
``...understanding the theoretical equations provided in the lab guide?" & 4.06\\ 
``...choosing an appropriate method for analyzing data (without explicit direction)?'' & 4.00\\ 
``...confiming previously known results?'' & 4.00\\ 
``...understanding the approximations and simplifications that are included in theoretical predictions?'' & 3.94\\ 
``...thinking about sources of systematic error?'' & 3.88\\ 
``...writing a lab report with the correct sections and formatting?'' & 3.81\\ 
``...overcoming difficulties without the instructor's help?'' & 3.69\\ 
``...designing and building things?" & 3.40\\ 
``...thinking up my own questions to investigate?" & 3.25\\ 
``...thinking about the purpose of the instructions in the lab guide?" & 3.13\\ 
``...reading scientific journal articles?'' & 2.87\\ 
``...randomly changing things to fix a problem with the experiment?'' & 2.00\\ 
\end{tabular}
\caption{The scale for the responses was (1) Unimportant, (2) Of Little Importance, (3) Moderately Important, (4) Important, (5) Very Important. }
\label{tab:GradeEmphasis}
\end{table*}

\section{Outcomes, Limitations, and Future Directions}

The Advanced Physics Lab course at CU was transformed using a process that involves three steps of (1) establishing learning goals, (2) aligning the course and curriculum to the goals, and (3) assessment.  A suite of lecture activities was developed, which covers data analysis in Mathematica and data acquisition in LabVIEW.  Also, a series of redesigned optics labs was modified in order to emphasize modeling and promote a deeper understanding of the apparatus and optical phenomena.  However, despite the satisfaction of creating a set of new materials, the greatest contribution from this transformation process is the continual refinement of the course through the iterative process of aligning learning goals, materials development, and assessments.  

Through assessment of the course, it became evident that there were limitations with the revised course.  The CU Advanced Lab is typical of many institutions in that the labs are mostly done in random order because we have limited experimental setups.  This lack of order in the sequence limits the amount of instructional scaffolding--meaning it is difficult to gradually build up experimental expertise because all of the labs should be about the same level of difficulty and sophistication.  Also, the lack of alignment with the canonical undergraduate lecture courses means that students are exposed to a lot of new physics content in addition to navigating a new experimental apparatus.  While it is commonly assumed that labs benefit students' learning of physics content, the reverse also is true, that a more sophisticated theoretical knowledge allows you to analyze experiments in a more sophisticated expert-like way.  For instance, a deeper theoretical knowledge can open up an analysis of the systematic error sources arising from the limitations of their models and measurement devices.  Lastly, certain parts of the course lacked adequate assessment tools that were linked with the learning goals, mainly the oral and written reports and final projects.

Future iterations of the course will respond to these limitations and other issues that came up through the assessments.  One short term priority is establishing and using rubrics for the creation and evaluation of lab notebooks, oral talks, written papers, and final projects.  Another short-term priority is modifying the activities and labs based on faculty and student feedback. On a longer timescale, when we consider how experimental reasoning is limited by students' lack of background in many areas of modern physics and optics, we believe a good strategy would be to develop sophisticated experiments that build as much as possible on students' prior course work in lecture courses.    

Finally, we plan to adapt some of our modern physics experiments to the revised learning goals and expand the work of upper-division lab transformation to the junior-level electronics lab course.

\section{Acknowledgments}
The authors would like to thank the CU Physics Department for contributing to the learning goals process.  We would also like to particularly thank faculty members Charles Rogers, Noel Clark, and Kevin Stenson for contributing their teaching insight and experimental expertise as they gave feedback on the previous and newly revised advanced lab course.  This work is supported by NSF CAREER PHY-0748742, NSF TUES DUE-1043028, JILA PFC PHY-0551010, the CU Science Education Initiative, and Integrating STEM Education at Colorado NSF DRL-0833364.  The views expressed in this paper do not necessarily reflect those of the National Science Foundation.

\bibliography{Publications-AJP_Lab_Transformation_MOD}

\end{document}